Article

# The emergence of globular clusters and globular-cluster-like dwarfs

Ethan D. Taylor[1 ✉], Justin I. Read[1], Matthew D. A. Orkney[2,3], Stacy Y. Kim[1,4], Andrew Pontzen[5], Oscar Agertz[6], Martin P. Rey[7], Eric P. Andersson[8], Michelle L. M. Collins[1] & Robert M. Yates[9]

Globular clusters (GCs) are among the oldest and densest stellar systems in the Universe, yet how they form remains a mystery[1]. Here we present a suite of cosmological simulations in which both dark-matter-free GCs and dark-matter-rich dwarf galaxies naturally emerge in the Standard Cosmology. We show that these objects inhabit distinct locations in the size–luminosity plane and that they have similar ages, age spread, metallicity and metallicity spread to globulars and dwarfs in the nearby Universe. About half of our simulated globulars form by means of regular star formation near the centres of their host dwarf, with the rest forming further out, triggered by mergers. The latter are more tidally isolated and more likely to survive to the present day. Finally, our simulations predict the existence of a new class of object that we call 'globular-cluster-like dwarfs' (GCDs). These form from a single, self-quenching, star-formation event in low-mass dark-matter halos at high redshift and have observational properties intermediate between globulars and dwarfs. We identify several dwarfs in our Galaxy, such as Reticulum II (refs. 2–4), that could be in this new class. If so, they promise unprecedented constraints on dark-matter models and new sites to search for metal-free stars.

GCs were first discovered by Abraham Ihle in 1665, yet there is still no consensus on how they form[1]. They are among the densest stellar systems known, with stellar masses $M_* \approx 10^{5-6}\,M_\odot$ and half-light radii $R_{1/2} \approx 1-10$ pc (ref. 1), making them an important source of gravitational waves[5] and possible sites for the formation of supermassive black holes seeds[6–8]. Most are assumed to have no dark matter, suggesting that they form along a distinct pathway to dark-matter-rich dwarf galaxies[9–11]. Current theories for GC formation fall into three main categories. The first suggests that they form in the same way as all star clusters but are simply the high-mass tail of the distribution[12]. The second suggests that they require special conditions to form, such as galaxy mergers[13,14], high-density converging gas flows[15,16], low-metallicity gas[17,18] or disk instabilities[19]. The third suggests that they form inside their own dark-matter halos[20–25]. It is hard for any one of these theories on its own to explain all of the GC observations, suggesting that they may act in together to some degree.

Low-mass dwarf galaxies, with stellar masses in a similar range to GCs ($M_* \approx 10^{3-7}\,M_\odot$) but much larger sizes ($R_{1/2} \approx 10-1{,}000$ pc), form at the same time as GCs in the early Universe[2]. Unlike GCs, dwarfs show clear evidence for dark matter from the kinematics of their stars and gas, extended star formation and large metallicity spreads. As such, GCs and dwarfs occupy distinct locations in the size–luminosity plane (see Fig. 1a) and have distinct metallicity and age distributions. However, this presents two puzzles. (1) How does the Universe conspire to form GCs and dwarfs at the same time in the early Universe? (2) And what are the objects in the region of parameter space in which GCs and dwarfs overlap?

In this paper, we set out to answer both questions using a suite of state-of-the-art simulations, Engineering Dwarfs at Galaxy formation's Edge (EDGE). EDGE models the smallest stellar systems to the present day in the Standard Cosmology at a spatial resolution of about 10 light-years (3 pc), at which we capture the impact of individual stellar supernovae on the surrounding interstellar medium[26,27] (Methods). This level of realism allows us to resolve the formation of galaxy-scale winds, driven by correlated star formation, which regulate the growth in stellar mass of the galaxy over time. We study 15 simulated EDGE galaxies with dark-matter halo masses in the range $10^7 < \frac{M_{200}}{M_\odot} < 10^{10}$, seven of which are centrals with halo mass $>10^9\,M_\odot$ and eight of which are satellites, selected to be tidally isolated (Methods), with halo mass in the range $10^7 < \frac{M_{200}}{M_\odot} < 10^9$. We extract bound stellar systems that survive to the present day (both dark-matter-rich dwarf galaxies and dark-matter-free star clusters) from these simulations using a new structure-finding algorithm that includes clustering in space, velocity and age, allowing us to confidently identify and track objects with >10 star particles (Methods).

[1]School of Mathematics and Physics, University of Surrey, Guildford, UK. [2]Institut de Ciencies del Cosmos (ICCUB), Universitat de Barcelona (IEEC-UB), Barcelona, Spain. [3]Institut d'Estudis Espacials de Catalunya (IEEC), Barcelona, Spain. [4]Carnegie Observatories, Pasadena, CA, USA. [5]Institute for Computational Cosmology, Department of Physics, Durham University, Durham, UK. [6]Lund Observatory, Division of Astrophysics, Department of Physics, Lund University, Lund, Sweden. [7]Department of Physics, University of Bath, Bath, UK. [8]Department of Astrophysics, American Museum of Natural History, New York, NY, USA. [9]Centre for Astrophysics Research (CAR), University of Hertfordshire, Hatfield, UK. ✉e-mail: e.d.taylor@surrey.ac.uk

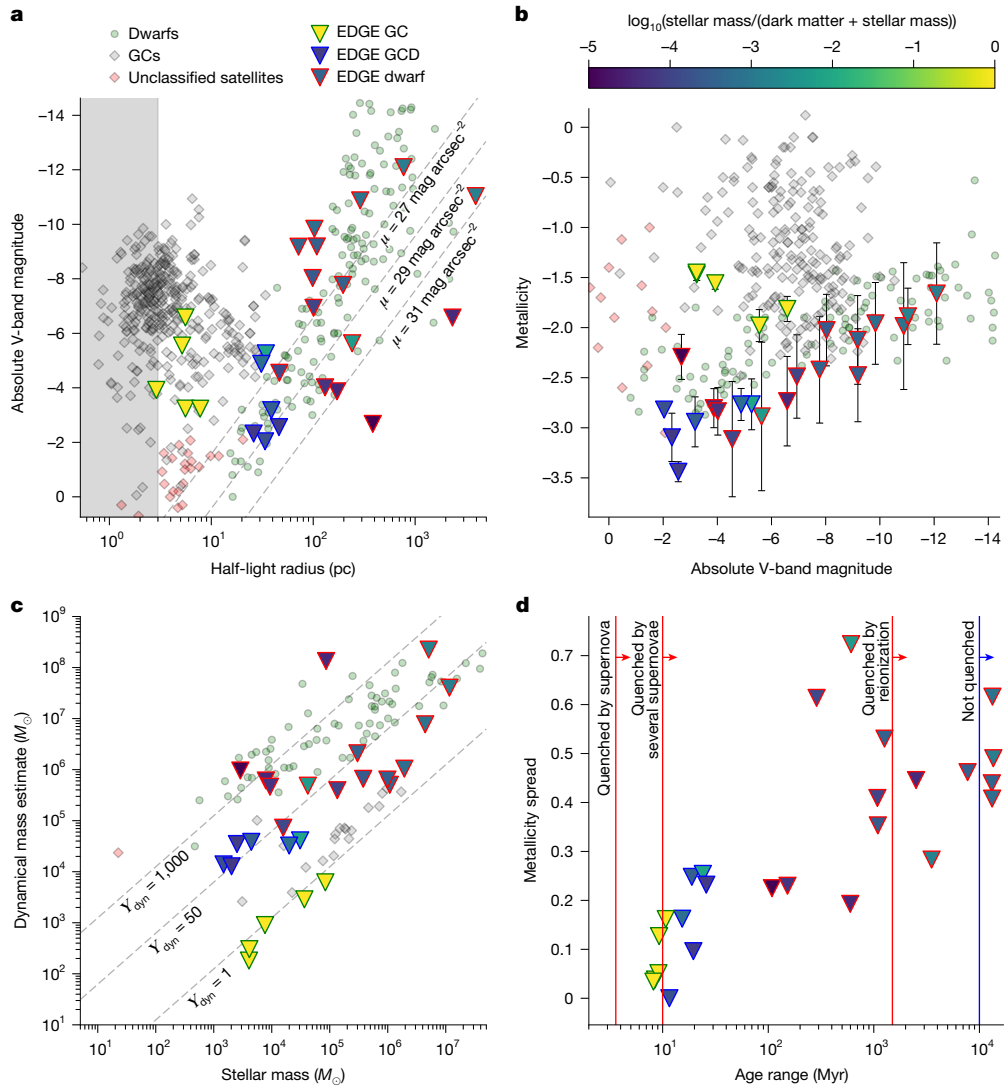

**Fig. 1 | Stellar systems formed in our EDGE simulations as compared with observational data for dwarf galaxies, GCs and unclassified satellites from the Local Group.** Observed GCs are marked by the grey diamonds, observed dwarfs by the green circles and unclassified satellites by the red diamonds. The EDGE objects (triangles) fall into three categories: GCs that are quenched by stellar winds and the first supernovae, have no dark matter and have little metallicity or age spread (triangles with a green border); dwarf galaxies that mostly comprise dark matter, with a high dynamical mass and large metallicity and age spread (triangles with a red border); and GCDs (triangles with a blue border) that are quenched by the first supernovae and have properties intermediate between GCs and dwarf galaxies. **a**, Absolute V-band magnitude ($M_V$) versus projected half-light radius. The EDGE spatial resolution (3 pc) is marked by the grey-shaded region. Lines of constant surface brightness are marked as dashed grey lines. The simulated EDGE stellar systems are coloured by their stellar-to-total mass fraction (see the colour bar). **b**, Metallicity versus $M_V$. The symbols are as in the key in **a**. The error bars show mass-weighted standard deviations. **c**, Dynamical mass within the half-light radius versus stellar mass. The symbols are as in the key in **a**. The dashed grey lines indicate constant dynamical-to-stellar mass ratios, as marked. **d**, Metallicity spread versus age spread. The symbols are as in the key in **a**. The vertical red lines mark the approximate times at which a single supernova, several supernovae and reionization quench star formation, as marked.

In Fig. 1, we show that, at the resolution of EDGE, a realistic population of both dwarf galaxies and GCs naturally emerge in our simulations. Defining 'metallicity spread' ($\sigma_{\rm [Fe/H]}$) as the standard deviation of [Fe/H] and 'age spread' ($\Delta_{\rm Age}$) as the difference in age between the oldest and the youngest stars in the stellar system, we find that our EDGE simulations produce three distinct types of object. The first are the GCs. These have their star formation shut down or 'quenched' by stellar winds and the first supernovae. They have no dark matter and little metallicity or age spread (triangles with a green border). The second are the dwarf galaxies. These mostly comprise dark matter, with a high dynamical mass (defined here using the mass estimator: $M_{\rm dyn} = R_{1/2} \sigma_{\rm los}^2 / G$, in which $R_{1/2}$ is the half-light radius, $\sigma_{\rm los}$ is the projected velocity dispersion and $G$ is Newton's gravitational constant[2]) and large metallicity and age spread (triangles with a red border). The third are a new type of object that we call GCDs (triangles with a blue border). The need for a third class is seen most clearly in Fig. 1c,d. GCDs form in their own dark-matter halos from a single self-quenching starburst, leading to them having an age spread ($\Delta_{\rm Age}$) and dynamical-to-stellar mass ratio ($Y$) in between that of GCs and dwarfs. Given this, we define GCDs as having $10 < \Delta_{\rm Age}$ (Myr) $< 50$ and $Y > 10$.

We find a close agreement between the observational properties of our simulated and real GCs. The EDGE GCs have size $R_{1/2} < 10$ pc, luminosity $-7 < M_V < -3$, metallicity $-2 < [{\rm Fe/H}] < -1.5$, dynamical-to-stellar mass ratio $Y \approx 1$, age spread $\Delta_{\rm Age} < 10$ Myr and metal spread $\sigma_{\rm [Fe/H]} < 0.2$ dex, all consistent with real GCs in the nearby Universe (see Fig. 1). Also, our GCs have a mean V−I colour of 0.98 with a standard deviation of 0.01, which is comparable with the observed 'blue' peak in the bimodal GC colour distribution (V−I = 0.95 ± 0.02 (ref. 28)). We do not capture, however, the formation of surviving GCs brighter than $M_V \approx −7$, with higher metallicity [Fe/H] > −1.5 and redder colours. These must form in higher-mass galaxies (for example, refs. 13,19), possibly by means of a distinct mechanism[29].

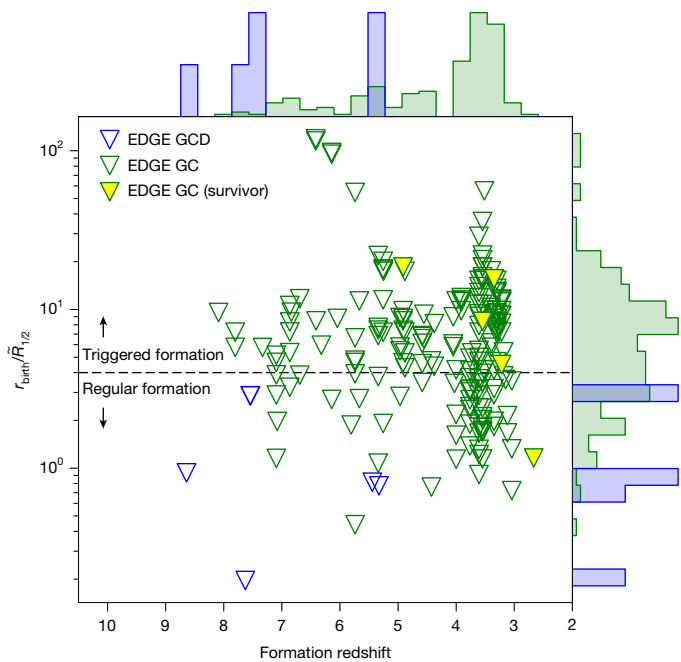

**Fig. 2 | GCs and GCDs in EDGE form over a wide range of birth radii and redshifts.** The GCs are marked by the triangles with a green border and the GCDs are marked by the triangles with a blue border. About half of the GCs form close to the centre of their host dwarf, with $r_{birth}/\widetilde{R}_{1/2} < 4$, in which $\widetilde{R}_{1/2}$ is a characteristic size scale for the host galaxy (see Methods). We define these as forming through 'regular star formation'. The other half form far from the centre of the host. We define these as 'triggered'. Only the five most massive GCs survive to the present day, four of which are triggered. This means that they form in a low-density environment that promotes their survival. GCDs form at higher redshift than the GCs and always close to the centre of their host dwarf.

Across all EDGE dwarfs, only five GCs survive to the present day. Four of these form in the most massive dwarf (with $M_{200} \approx 10^{10} M_\odot$), whereas one forms in the second most massive dwarf (with $M_{200} \approx 5.6 \times 10^9 M_\odot$). In Fig. 2, we show the birth radii of GCs in our EDGE simulations, $r_{birth}/\widetilde{R}_{1/2}$ (in which $\widetilde{R}_{1/2}$ is a characteristic size scale for the host galaxy), as a function of their formation redshift (see Methods). All except for one of our surviving GCs forms far from their host galaxy, placing them in a low-density environment that aids their survival. A further 184 GCs form in dwarfs of all halo masses over the redshift range $2.5 < z < 7.5$, peaking at a redshift of $z \approx 3$, but these have lower stellar mass and do not survive. We will study these in more detail in a forthcoming paper. (Note that these lower-mass GCs have too low density to fall to the centre of their host dwarf to form a central nuclear star cluster. However, nuclear star clusters can and do form in our EDGE simulations through a distinct mechanism[30]).

Our simulated EDGE dwarfs are similarly realistic, with a larger size at the same luminosity as the GCs ($R_{1/2} > 40$ pc), a strong correlation between [Fe/H] and luminosity (the mass–metallicity relation[31]), high dynamical mass $Y > 50$–100 and a broad age and metallicity spread.

Finally, the GCDs have properties intermediate between GCs and dwarfs, with sizes $R_{1/2} \approx 10$–60 pc, metallicities [Fe/H] < −2.75, dynamical-to-stellar mass ratio $Y \approx 50$ and age and metal spreads $\Delta_{Age} \approx 10$–20 Myr and $\sigma_{[Fe/H]} \approx 0.1$–0.3 dex, respectively. They form, like dwarfs, inside their own dark-matter halos and at systematically higher redshift than the GCs, peaking at $z \approx 8$ (see Fig. 2).

We now turn to the question of how GCs, dwarfs and GCDs form in our simulations. In Fig. 3, we show that GCs in EDGE form along two main pathways: through 'regular star formation' (top row) and merger-driven 'triggered star formation' (bottom row), encompassing the range of proposed formation mechanisms for GCs in the literature[12–14]. About half of the GCs form by means of regular star formation, which we define as having $r_{birth}/\widetilde{R}_{1/2} < 4$. The other half form further out, with the distribution extending to $r_{birth}/\widetilde{R}_{1/2} > 100$. All of the GCs form from pre-enriched gas such that their metallicity is about 1 dex higher than that of dwarfs at the same luminosity. The lack of any protective dark-matter halo means that the GCs self-quench by means of stellar winds or just one supernova and have, therefore, narrow age and metallicity spreads.

Our simulated EDGE dwarf galaxies form through regular star formation in low-mass dark-matter halos of virial mass $>10^9 M_\odot$. Their protective dark-matter halos are able to hold on to and recycle gas over many generations of star formation and their star formation must, therefore, be quenched by some external process. The lowest-mass dwarfs ($M_{200} < 2 \times 10^9 M_\odot$) are quenched by reionization (ionizing photons from galaxies and quasars[27]). More massive dwarfs ($M_{200} \approx 5 \times 10^9 M_\odot$) 'rejuvenate' their star formation after reionization, whereas the most massive dwarfs ($M_{200} > 10^{10} M_\odot$) form stars continuously to the present day[32].

Finally, GCDs form from a single star-formation event in the lowest-mass halos able to form stars before reionization. In our EDGE simulations, these have masses in the range $M_{200} \approx 10^{6.8-7.1} M_\odot$ at redshift $z = 5$–10, with a birth peak circular speed in the range $v_{max} \approx 7.2$–9.3 km s$^{-1}$. They are at the threshold density at which gas can cool and form stars; lower-density halos at the same redshift are starless, whereas more dense halos excite continuous star formation and host, therefore, dwarf galaxies. Half of the GCDs have their star formation initiated by the pollution of metals from a nearby stellar system. These have no 'metal-free' stars. The other half self-cool from pristine gas and have a high fraction (about 20%) of metal-free stars. If these GCDs can be found, they will be excellent sites to hunt for metal-free stars and to determine their impact on the next generation of stars. The presence of a dark-matter halo allows GCDs to form stars for longer than GCs and several supernovae are required to fully quench their star formation. This leads to a larger age and metal spread than the GCs. The dark-matter halo also raises the dynamical mass of the GCDs above that of the GCs but still roughly an order of magnitude lower than that of the dwarfs. We show an example GCD forming in Fig. 4.

Our EDGE simulations represent an important milestone in galaxy-formation theory. At a spatial resolution of about 10 light-years (3 pc), we find that both observationally realistic dark-matter-free GCs and dark-matter-rich dwarf galaxies naturally emerge in the Standard Cosmology. Previous work has captured the formation of star clusters in galaxy-formation simulations[14,33–35], whereas some studies have found that GC-like objects can form in dark-matter halos[21,22]. However, unique to EDGE is the simultaneous emergence of realistic GCs, dwarfs and a new class of object that we call GCDs.

GCDs promise unprecedented constraints on the nature of dark matter, as a thermal relic mass of about 10 keV is already sufficient to eliminate the halos in which GCDs form (for example, refs. 36,37). GCDs are also promising new sites to examine the physics of metal-free stars[38,39]. Given their potential importance, we may wonder whether GCDs could be hiding in plain sight in our cosmic backyard. The stellar stream C-19 has properties consistent with GCDs[40] and has already been proposed, from dynamical arguments, as a good candidate for a disrupting GC with a dark-matter halo[41]. Other promising candidates include Boötes V, Horologium I, Reticulum II, Boötes II, Draco II, Eridanus III and Delve I (refs. 2,42). If these are GCDs, we predict that they should have narrow age spreads

ArticleArticle

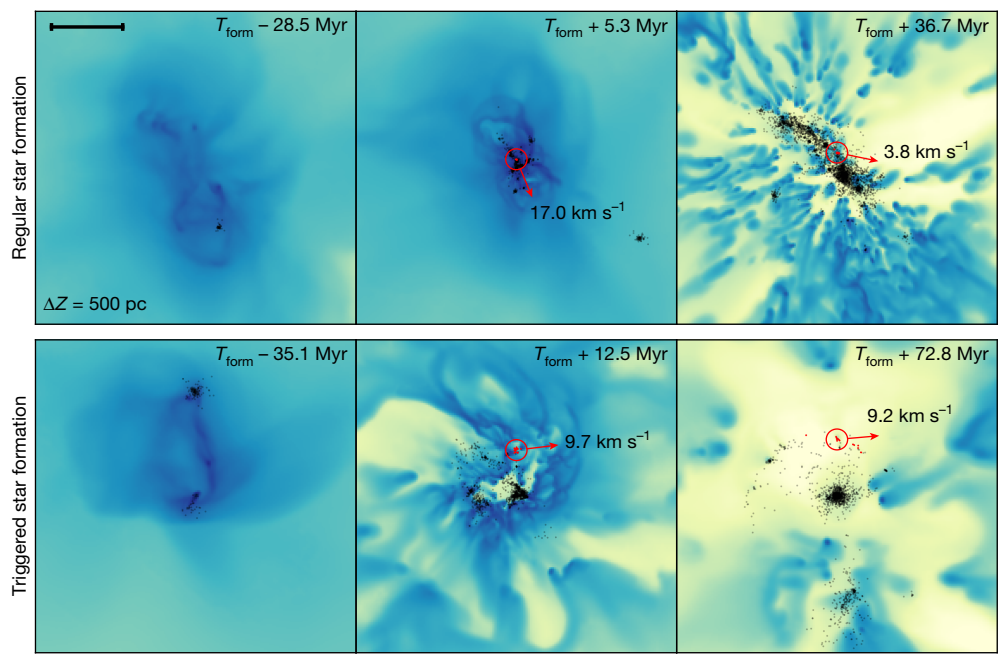

**Fig. 3 | Simulated EDGE GCs form in two main ways: through 'regular star formation' or merger-driven 'triggered star formation'.** The panels show the gas density of a slice of thickness 4 kpc along the line of sight just before (left), at (middle) and after (right) GC formation. The black points show the simulated star particles, with the forming GC highlighted with a red circle. The top row shows an example of GC formation through regular star formation and the bottom row through triggered star formation. Scale bar, 400 pc.

(10–20 Myr) and be older than ultra-faint dwarfs. Notably, Reticulum II shows evidence for such an ancient stellar population[3] and it also has unique chemistry, with $72^{+10}_{-12}\%$ of its stars enhanced in $r$-process elements[4]. This could be a signature of pollution from the very first generation of stars[43,44], consistent with the idea that it is a GCD. If so, the chemical imprint of metal-free stars may have already been found.

### Online content

Any methods, additional references, Nature Portfolio reporting summaries, source data, extended data, supplementary information, acknowledgements, peer review information; details of author contributions and competing interests; and statements of data and code availability are available at https://doi.org/10.1038/s41586-025-09494-x.

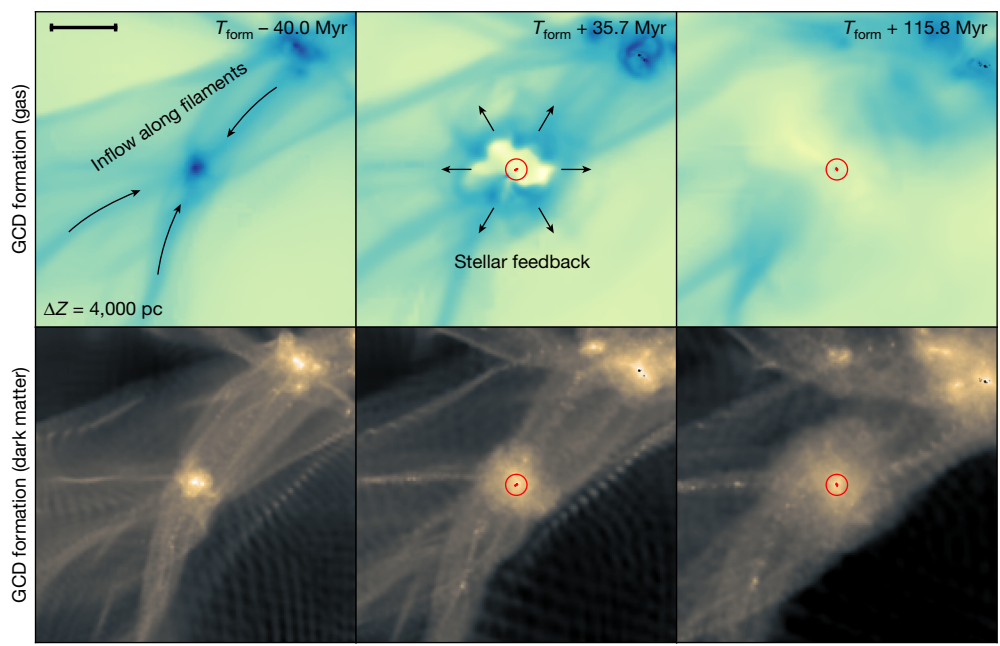

**Fig. 4 | A GCD forming in one of our EDGE simulations.** Gas cools in a low-mass dark-matter halo before reionization (left). Supernova feedback from the first star-formation event then self-quenches the GCD (middle) and no further star formation occurs (right). The points, labels and contours are as in Fig. 3 but now the bottom row shows contours of the dark-matter density. Scale bar, 800 pc.

# Article

## Methods

### Simulations

We use simulations with EDGE, a suite of hydrodynamical cosmological zoom-in simulations first described in ref. 27. Our suite includes five simulations of dwarfs with halo masses today (at redshift $z = 0$) in the range $10^9 < M_{200}/M_\odot < 3 \times 10^9$ that were first presented in refs. 27,32,45–47, one at a higher mass, $M_{200} = 5.6 \times 10^9 M_\odot$, that was first presented in ref. 30 and one at the highest mass, $M_{200} = 9.3 \times 10^9 M_\odot$, that we present for the first time in this work. We augment these dwarfs with a further eight dark-matter-rich objects that orbit within the high-resolution region of our EDGE simulations but are tidally isolated, with masses in the range $10^7 < M_{200}/M_\odot < 10^9$ and six objects that we classify as GCDs, with masses today in the range $10^{6.5} < M_{200}/M_\odot < 10^{7.5}$. We define 'tidally isolated' dwarfs as those that have not lost any mass to tides within 4 half-light radii. We report the properties of all of these objects in Extended Data Table 1. (Note that some of our EDGE halos have slightly different reported parameters (such as stellar mass, size and so on) as compared with the original reported values in refs. 27,47. This is because of the fact that these earlier works used the HOP halo finder[48] instead of the Amiga Halo Finder (AHF)[49], which we use here).

The initial conditions for our EDGE simulations were set up using the 'zoom-in' technique[50] to focus resolution on targeted halos within the initial dark-matter-only 50-Mpc$^3$ void region. This starts outs with a base resolution of $512^3$ particles and is evolved for a Hubble time. Isolated dark-matter-halos (see Extended Data Table 1) are then selected for resimulation at a higher resolution of 945 $M_\odot$ (fiducial resolution) or 117 $M_\odot$ (high resolution) per dark-matter particle. All simulations use cosmological parameters $\Omega_\Lambda = 0.691$, $\Omega_b = 0.045$, $\Omega_m = 0.309$ and $H_0 = 67.77$ km s$^{-1}$ Mpc$^{-1}$ taken from ref. 51. We evolve the simulations to the present day using the adaptive mesh refinement simulation code RAMSES[52], with a base grid resolution for the gas hydrodynamics of 3 pc. We refer to this base grid resolution throughout this paper as our 'spatial resolution'. This is because in grid codes such as RAMSES, stellar and dark-matter particle orbits are well recovered even down to separations approaching this resolution limit[53]. To reduce the effects of numerical diffusion, the velocity of each zoomed halo is adjusted to minimize its motion through the simulation box[54]. The initial conditions for the simulations are set up using the GenetIC software[55]. This allows us to 'genetically modify' individual galaxies to forensically investigate the impact of different mass accretion histories on galaxy properties while keeping its present-day mass and environment unaltered[56,57].

We model gas cooling, star formation and stellar feedback as described in ref. 27. Gas is allowed to cool to $T < 100$ K using the fine-structure cooling rates from ref. 58. Stars are formed in a given cell if the gas temperature $T_{gas} < 100$ K and density $\rho_{gas} > 300$ m$_H$ cm$^{-3}$, with a formation rate given by the Schmidt law: $\dot{\rho}_* = \varepsilon_{ff} \rho_{gas}/t_{ff}$, in which $t_{ff} = \sqrt{3\pi/(32G\rho)}$ is the local gas freefall time and $\varepsilon_{ff} = 0.1$ is the star-formation efficiency. In both the fiducial and high-resolution simulations, stars are modelled as approximately 300-$M_\odot$ particles that represent a single stellar population with a Kroupa initial mass function. We model heating from reionization through a spatially uniform, time-dependent ultraviolet background, as implemented in the public RAMSES version[32]. Stellar feedback is implemented as in ref. 59 and includes asymptotic giant branch winds, radiation pressure from young stars and Type I and Type II supernovae explosions, modelled as discrete thermal injection events. We do not model the spatial and dynamical distribution of individual stars nor the impact of 'runaway' stars (for example, refs. 60–62). We will explore this in future work[63]. At the resolution of our EDGE simulations, we capture the impact of almost all individual supernovae on their surrounding interstellar medium, without the need for delayed cooling, further momentum injection or similar (for example, refs. 26,64,65). This substantially increases the robustness and predictive power of the simulation results (see also, for example, refs. 27,66–68). We track iron and oxygen abundances separately but do not follow any other elements. (We will study the chemistry of our simulated objects in more detail in future work using the EDGE2.0 simulations that track eight different elements[69]). In previous EDGE papers, we have explored the effects of radiative transfer[27,69], changing our background reionization model[32] and changing the initial mass function of stars[70]. These choices can alter the stellar masses of our dwarfs by a factor of approximately 2 but do not otherwise alter the main results and conclusions presented here.

Despite the high spatial and mass resolution of our EDGE simulations (3 pc and see Extended Data Table 1), we may worry that we do not correctly resolve the smallest star clusters and dwarfs. We already explored the impact of dark matter mass resolution on dwarf galaxies in EDGE in ref. 71 (Appendix A), showing that our dark-matter halo density profiles are well converged on size scales sufficient to resolve the half-light radii of the dark-matter-rich objects presented in this paper. To test the impact of our spatial resolution on our star clusters, we run Halo605 at 2× (1.5 pc) and 4× (0.75 pc) higher spatial resolution down to redshift $z = 5$. The results are shown in Extended Data Fig. 1. Note that both the mass and the metallicity of our star clusters are well converged across all three simulations. The size of our simulated star clusters, however, shrinks with increasing resolution. This means that we should treat the birth sizes of our star clusters as upper bounds. This does not affect, however, our key results and conclusions for two main reasons. First, even if our simulated star clusters have smaller sizes, it remains the case that they naturally separate from dwarfs in the size–luminosity plane, they have distinct metallicity distributions from dwarfs and that we predict the existence of a new class of object: GCDs. Second, we do not model collisional two-body stellar dynamics in this work (more on this below). Such two-body effects cause star clusters to expand to a size in equilibrium with their local tidal field (for example, ref. 72), erasing memory of their birth sizes at the present day.

Finally, we may worry that the GCs that form in our EDGE simulations are collisional stellar systems, meaning that the individual gravitational encounters between stars are important for their evolution (for example, ref. 72). We explicitly model this in a companion paper (Taylor et al. in preparation) in which we resimulate each EDGE GC from birth to the present day using the direct N-body code Nbody6DF, similarly to in ref. 73. There we show that EDGE slightly overestimates the rate of tidal destruction of GCs once two-body effects are taken into account. This is because our EDGE simulations: (1) underestimate the true density of our GCs at birth (see Extended Data Fig. 1) and (2) cause our GCs to artificially expand owing to insufficient force resolution. This means that the results we present here should be taken as a lower bound on the number of surviving GCs and an upper bound on their sizes.

### Structure finding

To search for structures and sub-structures in the EDGE simulation volume, we first use the AHF[49]. We include all gravitationally bound objects (hereafter 'halos') with a minimum of 100 particles (dark matter and/or stars)[74]. AHF has been well tested on mock simulation data and is a widely used community tool[75,76]. However, we found that it was unable to detect dark-matter-free star clusters or the smallest bound dark-matter-rich structures in EDGE — both of which were discoverable by visual inspection of the simulation output (AHF alone missed about 15% of our GCDs and about 90% of our GCs). This difficulty with structure finding in deep cosmological hydrodynamic zooms is a known problem (for example, refs. 76,77). To solve it, we augment AHF with the multidimensional cluster finder, HDBScan[78]. This includes both velocity and (for stars) age dimensions in the clustering analysis, allowing us to reliably detect much smaller structures (>10 particles) than can be found with AHF alone. Our method proceeds as follows. First, we run HDBScan using positional data (centred on the main halo in the simulation), velocity data (momentum-centred on the main halo) and the dimensionless standardized 'Z-score' of the birth times based on the main halo: $Z_i = (\tau_i - \bar{\tau})/\sigma_\tau$, in which $\tau_i$ is the $i$th star particle's birth time and $\bar{t}$ and $\sigma_t$ are the mean and standard deviation of all birth times, respectively. We use units of parsecs for the distances, km s$^{-1}$ for the velocities and the dimensionless Z-score, thereby placing each of our clustering data dimensions on a size scale appropriate for star clusters. HDBScan reports a probability that each particle is associated with a given group. Next, we group particles with their highest-ranking group, assuming that the membership probability is greater than 75%. We treat these groups as 'seeds' around which we hunt for further group members and/or prune misidentified members. To do this, we first centre on each of the groups in position and momentum. Then we fit a Plummer sphere[79] to the star particles belonging to that group at present. We expand the group membership using the closest 75 star and 75 dark-matter particles to the group centre. This allows us to test whether any nearby dark matter particles are also bound to the group and to make sure that we have not missed any nearby star particles (our results are not sensitive to this choice of 75 neighbours). In this initial group-growth step, we exclude any particle that is outside 3 standard deviations from the mean of the original group in velocity and Z-score. Next, the stars and dark matter in this now-expanded group have their velocity magnitudes compared with the escape velocity of their fitted Plummer sphere. Any particles that exceed this escape velocity are unbound and are, therefore, removed from the group. This process is iterated, with a new Plummer sphere fit at each iteration and further particles removed until the group converges, leaving a gravitationally bound group of stars and/or dark matter. (Note that we ignore gas in this analysis because in all cases, we find that the gas contributes negligibly to the potential once star clusters and/or dwarf galaxies have formed). We merge our new GC groups, found as described above, with the AHF halo catalogue, avoiding any double counting between the two. Finally, each group is traced back to its birth snapshot and manually inspected to confirm that it is a genuine bound object. Highly extended objects—probably unbound owing to tidal forces—and/or other spurious structures are rejected in this final step.

### Post-processing of simulation output

We use the packages PyNbody[80] and Tangos[81] to analyse our simulation output. For our simulated GCs, dwarfs and GCDs, we calculate V-band and I-band luminosities as follows. First, we take each star particle and treat it as a single stellar population with fixed age and metallicity. We use the Padova stellar evolution library[82,83] to calculate the luminosity over a grid of ages and metallicities and interpolate over this for each particle. The final luminosity in a given waveband is then the sum of all of the contributing star particles.

When discussing the birth radii of our simulated GCs, we normalize these to a characteristic size for the host dwarf $\widetilde{R}_{1/2} = 0.015 r_{200}$, in which $r_{200}$ is the virial radius of the host (see ref. 84). We use $\widetilde{R}_{1/2}$ instead of their actual half-light radius because at early times, when dwarfs undergo many mergers and star clusters are forming, the half-light radius can fluctuate substantially from output to output.

Because, to a good approximation, our star clusters are single stellar populations, we define their 'formation redshift' as corresponding to the mean age of their stars.

### Observational data sample

The observational data points overplotted on Fig. 1 were taken from the Local Volume Database (https://github.com/apace7/local_volume_database)[85], which compiles unclassified satellite data from refs. 86–108, dwarf galaxies from refs. 109–144 and GCs from refs. 145–157.

### Data availability

Raw simulation data can be made available from the authors on request, including software to load in and analyse the data. Files to regenerate the initial conditions of the simulations used in the paper have been uploaded at https://zenodo.org/records/16536387 (ref. 158). The DOI associated with this dataset is https://doi.org/10.5281/zenodo.16536387.

### Code availability

All software used to set up, run and analyse the simulations presented is open source and publicly available at the following links: https://github.com/ramses-organisation/ramses; http://popia.ft.uam.es/AHF/; https://hdbscan.readthedocs.io/; https://pynbody.github.io/; and https://github.com/pynbody/genetIC. The following GitHub repository contains all of the scripts and data required to reproduce the figures in this paper: https://github.com/EthTay/EDGE-GC.

**Acknowledgements** E.D.T. acknowledges support from the UKRI Science and Technology Facilities Council (STFC; grants ST/V50712X/1 and ST/Y002865/1). J.I.R. and M.L.M.C. acknowledge support from STFC grants ST/Y002865/1 and ST/Y002857/1. M.D.A.O acknowledges funding from the European Research Council (ERC) under the European Union's Horizon 2020 research and innovation programme (grant agreement No. 852839). O.A. acknowledges support from the Knut and Alice Wallenberg Foundation, the Swedish Research Council (grant 2019-04659) and the Swedish National Space Agency (SNSA Dnr 2023-00164). E.P.A. acknowledges support from US NSF grants AST18-15461 and AST23-07950 and NASA ATP grant 80NSSC24K0935. A.P. received support from the European Union's Horizon 2020 research and innovation programme under grant agreement no. 818085 GMGalaxies. This work used both the DiRAC Data Intensive service (DIaL2) hosted and managed by the University of Leicester Research Computing Service and the DiRAC@Durham (cosma6) facility hosted and managed by the Institute for Computational Cosmology on behalf of the STFC DiRAC HPC Facility (www.dirac.ac.uk). The equipment was financed by BEIS capital funding through STFC capital grants ST/P002293/1, ST/R002371/1 and ST/S002502/1, Durham University and STFC operations grant ST/R000832/1. DiRAC is part of the National e-Infrastructure. We wish to thank A. Fattahi for helpful comments on an early draft of this work.

**Author contributions** E.D.T. is the lead author. He led the running and analysis of the higher-mass simulations and the convergence tests. He developed the structure-finding analysis pipeline. He wrote the first draft of the manuscript. J.I.R. is PI of the EDGE project. He co-conceived the idea for this project and led the proposal to win computer time on the UK national supercomputer, DiRAC. He contributed to writing the paper and assisted with data analysis and interpretation of the results. M.D.A.O. was involved in the running and management of lower-mass EDGE simulations. He assisted with the research, data analysis and interpretation of the results. He also worked on the figure-plotting scripts, assisting with all figures in the paper. He helped to review and edit the manuscript. S.K. assisted with data analysis and the interpretation of the results and reviewed and provided feedback on the manuscript. A.P. co-conceived the EDGE project and simulations. He provided methodological and software pipeline development for both the initial conditions and analysis. He reviewed and provided feedback on the manuscript. O.A. co-conceived the EDGE project. He provided methodological and software development for the EDGE sub-grid physics modules in the RAMSES code. He reviewed and provided feedback on the manuscript. M.P.R. produced the initial conditions for the simulations and contributed to the data analysis and curation pipeline. E.P.A. contributed to the development of the sub-grid physics modules in RAMSES. He contributed to the interpretation of the results, assisted with setting up and running the tests for the convergence study and reviewed and provided feedback on the manuscript. R.M.Y. contributed to the interpretation of the results and writing of the manuscript. M.L.M.C. assisted in the data analysis and the interpretation of the results. She provided feedback on the results and their comparisons with observations.

**Competing interests** The authors declare no competing interests.

**Additional information**
**Correspondence and requests for materials** should be addressed to Ethan D. Taylor.
**Peer review information** *Nature* thanks Carmela Lardo and the other, anonymous, reviewer(s) for their contribution to the peer review of this work.
**Reprints and permissions information** is available at http://www.nature.com/reprints.

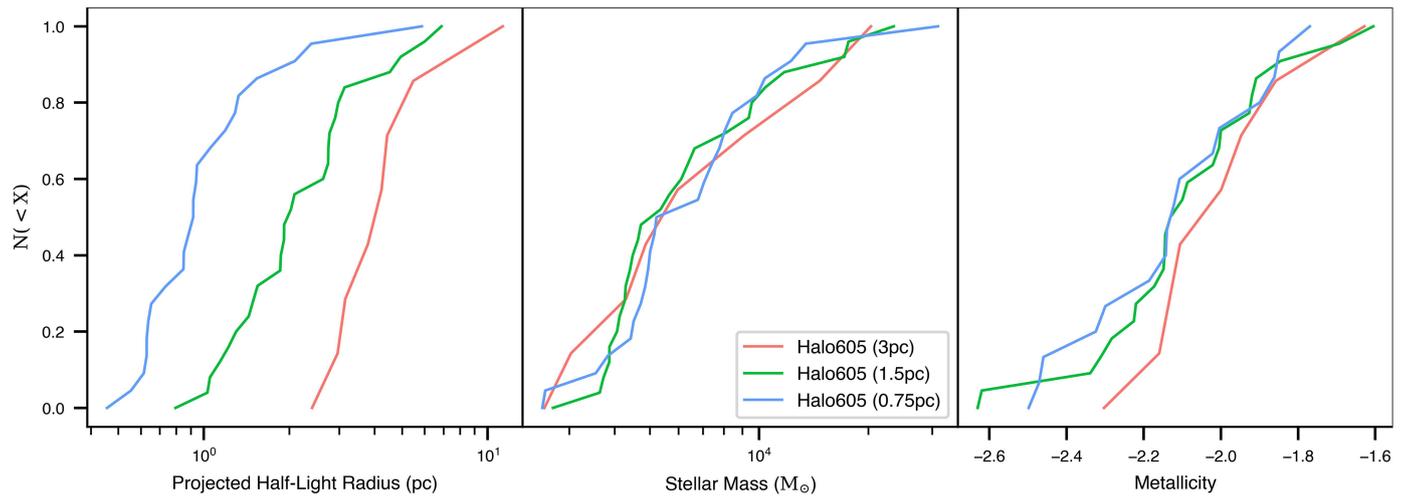

**Extended Data Fig. 1 | Testing the numerical convergence of simulated GCs in EDGE.** To test the impact of our spatial resolution on our results, we reran Halo605 (red) down to a redshift, $z = 5$, at 2× (green) and 4× (blue) higher spatial resolution. The left, middle and right panels show the cumulative distribution functions of the size (left), stellar mass (middle) and metallicity (right) of the simulated GCs in each simulation. Notice that both the mass and metallicity of our GCs are well converged across all three simulations. The size of our simulated GCs, however, shrinks with increasing resolution. This means that we should treat the birth sizes of our GCs as upper bounds. This does not affect, however, the main results and conclusions of our work (see text for details).

**Extended Data Table 1 | Properties of the EDGE dwarf galaxies at redshift z = 0 used in this paper**

| Label | Resolution $[m_{dm}, m_{gas}, m_*]/M_\odot$ | $\mathrm{Log}_{10}(M_{200}/M_\odot)$ | $r_{200}$ kpc | $\mathrm{Log}_{10}(M_*/M_\odot)$ | $M_V$ | $R_{1/2}$ pc | [Fe/H] dex | Type |
|---|---|---|---|---|---|---|---|---|
| Halo1445 | [117,18,300] | 9.11 | 23.1 | 5.15 | -6.9 | 100.8 | -2.5 | Cen |
| Halo1459 | [117,18,300] | 9.15 | 23.7 | 5.58 | -8.0 | 98.8 | -2.0 | Cen |
| Halo624 | [117,18,300] | 9.42 | 29.1 | 6.04 | -9.2 | 71.8 | -2.1 | Cen |
| Halo605 | [117,18,300] | 9.5 | 31.1 | 6.28 | -9.8 | 102.8 | -2.0 | Cen |
| Halo600 | [117,18,300] | 9.5 | 31.2 | 5.99 | -9.2 | 108.5 | -2.5 | Cen |
| Halo383early | [939,161,300] | 9.75 | 37.6 | 6.64 | -10.9 | 289.4 | -2.0 | Cen |
| Halo383Massive | [939,161,300] | 9.97 | 44.6 | 7.03 | -12.1 | 771.7 | -1.7 | Cen |
| Halo1459a | [117,18,300] | 8.55 | 14.9 | 3.5 | -2.7 | 384.5 | -2.3 | Sat |
| Halo1459b | [117,18,300] | 8.11 | 10.6 | 4.2 | -4.6 | 46.6 | -3.1 | Sat |
| Halo624a | [117,18,300] | 8.90 | 17.2 | 5.5 | -7.8 | 196.5 | -2.4 | Sat |
| Halo383earlya | [939,161,300] | 9.34 | 27.5 | 4.9 | -6.6 | 2326 | -2.7 | Sat |
| Halo383earlyb | [939,161,300] | 8.31 | 12.4 | 3.9 | -3.9 | 171.5 | -2.8 | Sat |
| Halo383earlyc | [939,161,300] | 8.22 | 11.6 | 3.95 | -4.0 | 130.8 | -2.8 | Sat |
| Halo383Massivea | [939,161,300] | 9.52 | 31.3 | 6.70 | -11.0 | 3928 | -1.9 | Sat |
| Halo383Massiveb | [939,161,300] | 6.89 | 3.8 | 4.60 | -5.6 | 241.2 | -2.9 | Sat |
| Halo624-GCDa | [117,18,300] | 7.51 | 6.7 | 3.5 | -2.6 | 46.2 | -3.4 | GCD |
| Halo624-GCDb | [117,18,300] | 7.16 | 5.1 | 3.6 | -3.2 | 38.8 | -2.9 | GCD |
| Halo624-GCDc | [117,18,300] | 6.93 | 4.3 | 3.2 | -2.0 | 33.7 | -2.8 | GCD |
| Halo624-GCDd | [117,18,300] | 6.53 | 1.5 | 4.48 | -5.3 | 34.3 | -2.8 | GCD |
| Halo1459-GCDa | [117,18,300] | 7.46 | 6.5 | 3.30 | -2.3 | 26.0 | -3.1 | GCD |
| Halo1445-GCDa | [117,18,300] | 7.47 | 6.5 | 4.30 | -4.9 | 31.0 | -2.8 | GCD |

From left to right, the columns show: the simulation label; the mass resolution (dark matter/gas/stars); the halo virial mass; the halo virial radius; the stellar mass; the absolute V-band magnitude; the projected half-light radius; the stellar mass-weighted average iron abundance; and whether the object is a central dwarf (Cen), satellite (Sat) dwarf or a GCD. Units for each column are given in the second row.